\documentclass[11pt,a4paper]{article}
\pdfoutput=1

\usepackage[]{Packages}

\bibliography{GaugeBetheRef}

\usepackage{ReviewPackages}

\allowdisplaybreaks[1]

\def\tbl{\caption}
\def\botrule{\bottomrule}
\def\colrule{\midrule}

\preprint{\texttt{CERN-PH-TH/2013-231}}

\newcommand{\OfficialTitle}{Deformed supersymmetric gauge theories from the fluxtrap background}

\title{\vspace{2cm}
  {\huge   \textbf{\dosserif\OfficialTitle}}
}

\hypersetup{pdfauthor={Domenico Orlando and Susanne Reffert},pdftitle={\OfficialTitle}}

\author{%
  \begin{minipage}{.8\linewidth}
    \vspace{1cm}
    \begin{center}
      {\small \textbf{Domenico Orlando} and \textbf{Susanne Reffert}}
    \end{center}
    \vspace{1cm}
    \begin{minipage}{\linewidth}\centering
      {\itshape \footnotesize
        Theory Group, Physics Department, \\ Organisation européenne pour la recherche nucléaire (CERN) \\ CH-1211 Geneva 23, Switzerland
      }
    \end{minipage}
  \end{minipage}
}

\date{}

\begin{document}

\setstretch{1.1}

\numberwithin{equation}{section}

\begin{titlepage}

  \maketitle

  \thispagestyle{empty}

  \vfill
  \abstract
{The fluxtrap background of string theory provides a transparent and algorithmic way of constructing supersymmetric gauge theories with both mass and Omega-type deformations in various dimensions. In this article, we review a number of deformed supersymmetric gauge theories in two and four dimensions which can be obtained via the fluxtrap background from string or M--theory. Such theories, the most well-known being Omega--deformed super Yang--Mills theory in four dimensions, have met with a lot of interest in the recent literature. The string theory treatment offers many new avenues of analysis and applications, such as for example the study of the gravity duals for deformed $\mathcal{N}=4$ gauge theories.}
\vfill

\end{titlepage}



\section{Introduction}

\begin{figure}
\centerline{\includegraphics[width=1\textwidth]{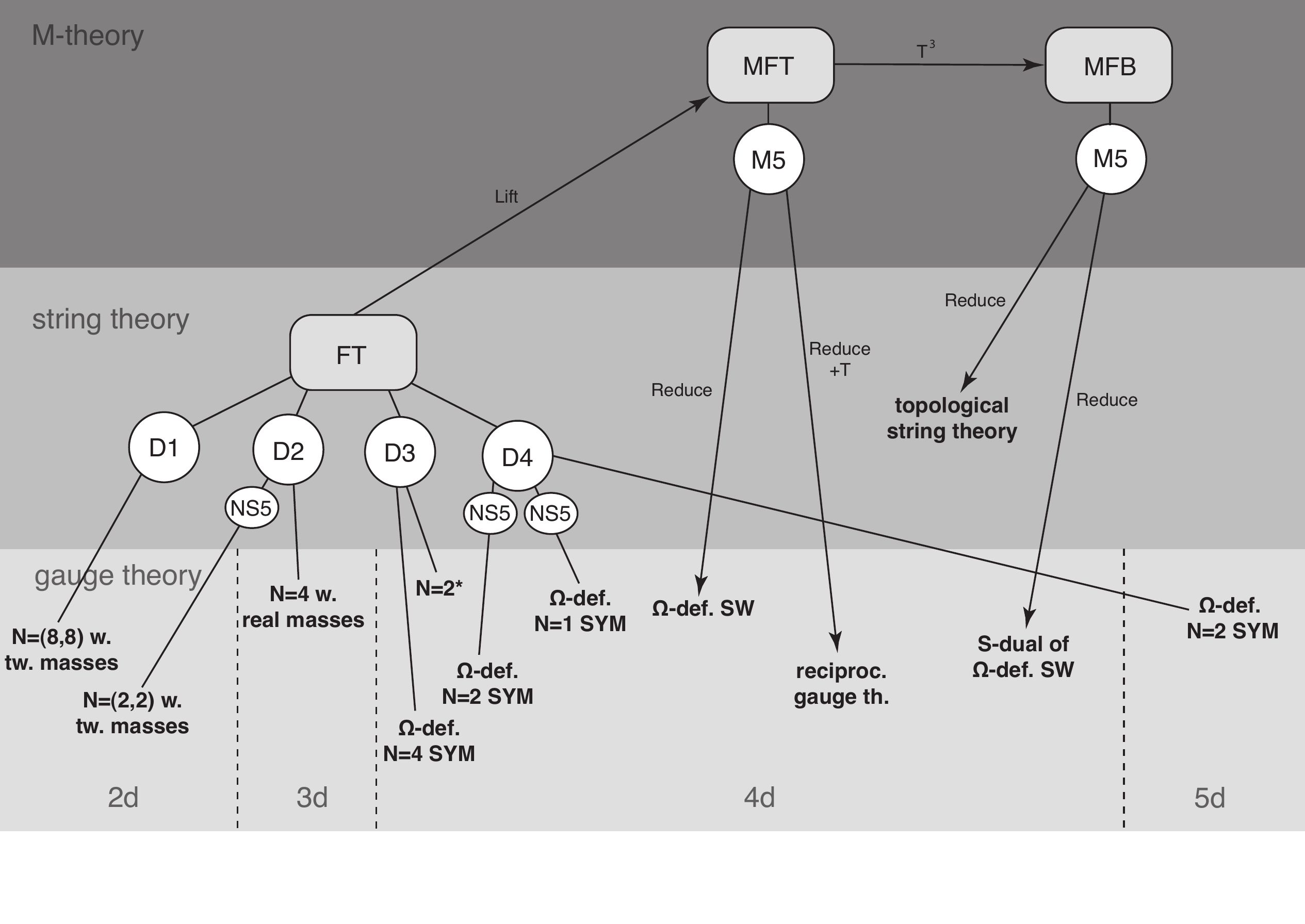}}
\vspace*{8pt}
\caption{An overview map of deformed supersymmetric gauge theories obtained from the string-- and M--theory fluxtrap backgrounds\label{fig:map}}
\end{figure}

In recent years, deformed supersymmetric gauge theories have played a prominent role in theoretical physics. The best-known examples involve $\Omega$--deformed $\mathcal{N}=2$ \ac{sym} theory which appears in the contexts of instanton localization~\cite{Nekrasov:2002qd, Nekrasov:2003rj, Pestun:2007rz}, topological string theory~\cite{Hollowood:2003cv, Iqbal:2007ii, Antoniadis:2010iq, Krefl:2010fm,Huang:2010kf, Aganagic:2011mi, Antoniadis:2013bja}, the \ac{agt} correspondence~\cite{Gaiotto:2009we, Alday:2009aq} and the gauge/Bethe correspondence~\cite{Nekrasov:2009rc}.  Other interesting examples, also in relation with integrability, involve two- and three-dimensional gauge theories with twisted masses~\cite{Nekrasov:2009uh, Nekrasov:2009ui}. The authors have shown in a series of papers that these deformed supersymmetric gauge theories have a \emph{common string theory realization}~\cite{Hellerman:2011mv, Reffert:2011dp, Orlando:2011nc, Hellerman:2012zf, Hellerman:2012rd, Lambert:2013lxa} and can thus be analyzed via string theoretic methods.
The great strength of this approach is that it makes manifest the fact that different kinds of gauge theory deformations which were thought to be unrelated have the \emph{same} origin in string theory.


\bigskip
In this article, we review the various deformed supersymmetric gauge theories that can be obtained by placing different brane set-ups into the so-called \emph{fluxtrap} background or its M--theory lift in a unified manner. Among them are $\Omega$--deformed \ac{sym} in four dimensions with $\mathcal{N}=4$, $\mathcal{N}=2$  and $\mathcal{N}=1$ supersymmetry, as well as two-dimensional gauge theories with twisted mass deformations and three-dimensional gauge theories with real mass deformations, to name just a few examples. Figure~\ref{fig:map} shows an (non-exhaustive) overview over the different gauge theories that can be constructed from the fluxtrap background, most of which will be discussed at least briefly in the following.

This unified string theory framework 
lends a geometric interpretation to a variety of gauge theoretic phenomena such as instanton localization.

\bigskip

The type of gauge theory deformation resulting from the fluxtrap background depends on how the D--branes are placed into the fluxtrap with respect to the deformations in the bulk. There are basically two possibilities, which can be combined.
The background monodromies being \emph{orthogonal} from the brane world-volume give rises to \emph{mass-type} deformations for the scalar fields encoding brane fluctuations in the deformed directions\footnote{Deformed directions away from the brane world-volume without an associated scalar field result in R--symmetries for the gauge theory.}, see Table~\ref{tab:off}.
When the background deformation happens \emph{on} the brane world-volume, the effective gauge theory receives an \emph{$\Omega$--type} deformation where Lorentz invariance is broken, see Table~\ref{tab:on}.
Of course it is possible to construct also gauge theories with both types of deformation present.

\begin{table}
\centering
  \begin{tabular}{lcccccccccc}
    \toprule
     fluxtrap &  & & & & \ep{\epsilon_i} &  \ep{\epsilon_j}  & &        \\
    D--brane &  &\X & \X &\X  & \ep{\phi_i}  & & & \\
        \botrule
  \end{tabular}
    \label{tab:off}
  \tbl{D--brane configuration in the fluxtrap corresponding to a twisted mass \( \epsilon_i \) for the field \( \phi_i \).}
\end{table}
\begin{table}
  \centering
 {
\begin{tabular}{lcccccccccc}
    \toprule
     fluxtrap &  & & \ep{\epsilon_i} &  \ep{\epsilon_j}  &   & &     \\
    D--brane & &  &\X & \X &\X  & \X & &   & \\
        \botrule
  \end{tabular}
    \label{tab:on}}
  \tbl{D--brane configuration in the fluxtrap corresponding to a \( \Omega \)--deformed gauge theory. }
\end{table}

\bigskip

The concrete advantages of our construction are:
\begin{itemize}
\item it provides an algorithmic way of generating new deformed gauge theories, such as \( \Omega \)--deformed \( \mathcal{N} = 1 \) \ac{sym};
\item it leads to a new way to describe the effective low-energy dynamics of the deformed theories via M--theory;
\item it gives a direct way of constructing the gravity duals to the deformed \( \mathcal{N} = 4 \) theories.
\end{itemize}

The plan of this article is the following. In Section~\ref{sec:bulk}, the background deformation is introduced in string theory. The fluxbrane and fluxtrap are discussed both in string (Sec.~\ref{sec:FB} and \ref{sec:FT}) and M--theory (Sec.~\ref{sec:MFT} and \ref{sec:MFB}), as well as the S--dual version which leads to an RR fluxtrap (Sec.~\ref{sec:RRFT}). Also the so-called reciprocal frame is derived (Sec.~\ref{sec:rec}). The supersymmetries of the deformed background are discussed in Section~\ref{sec:susy}. 

Our discussion of deformed supersymmetric effective gauge theories starts out in Section~\ref{sec:2d} with two two-dimensional examples, \( \Omega \)--deformed $\mathcal{N}=(8,8)$ theory with twisted masses (Sec.~\ref{sec:N88}), and $\mathcal{N}=(2,2)$ theory with twisted masses (Sec.~\ref{sec:N22}).
Section~\ref{sec:4d}, which treats deformed effective gauge theories in four dimensions kicks off with $\Omega$--deformed $\mathcal{N}=4$  \ac{sym} and $\mathcal{N}=2^*$ theory (Sec.~\ref{sec:N4}).
The archetypical example for the class of deformed supersymmetric gauge theories discussed in this article, $\Omega$--deformed $\mathcal{N}=2$  \ac{sym}, is discussed in Section~\ref{sec:N2}.
It is also possible to construct $\Omega$--deformed $\mathcal{N}=1$ theory from a modified brane set-up involving non-parallel NS5--branes (Sec.~\ref{sec:N1}). 
In Section~\ref{sec:SW}, a more complicated example which takes the route via the M--theory fluxbrane background is discussed, namely the derivation  of the $\Omega$--deformed \ac{sw} Lagrangian.
The reciprocal gauge theory which bears some striking similarities to Liouville theory is discussed in Section~\ref{sec:recg}. Having obtained deformed $\mathcal{N}=4$ \ac{sym} theories in Sec.~\ref{sec:N4}, we finally study their Polchinski--Strassler-type gravity duals in Section~\ref{sec:AdSCFT}. We close in Section~\ref{sec:conc} with some conclusions.

\section{The bulk deformation}\label{sec:bulk}

In this section, we will introduce the string theory bulk deformation, which will give rise to the gauge theory deformations we will be discussing in the following. The deformation can take place either in \tIIA or \tIIB string theory. In oder to describe the deformed string theory background, we divide ten dimensional Euclidean space into four planes each parameterized by a radial coordinate $\rho_i$ and an angular coordinate $\theta_i$, while the $x_8,\, x_9$--directions form a torus $T^2$, see Table~\ref{tab:bulk}. Each of the four planes can in principle  be deformed via a deformation parameter $\epsilon_i$.
\begin{table}[h]
  \centering
 { \begin{tabular}{lcccccccccc}
    \toprule
    \( x \)   & 0               & 1               & 2               & 3  & 4  & 5  & 6  & 7 & 8  & 9  \\
     & \ep{(\rho_1,\theta_1)} &  \ep{(\rho_2,\theta_2)}  & \ep{(\rho_3,\theta_3)} & \ep{(\rho_4, \theta_4)} & \ep{v} \\
    \colrule
    fluxbrane & \ep{\epsilon_1} & \ep{\epsilon_2} & \ep{\epsilon_3} &  \ep{\epsilon_4} & \T & \T               \\
        \botrule
  \end{tabular}
  \label{tab:bulk}}
    \tbl{Coordinate names and $\epsilon$--deformed directions. Circles denote periodic Melvin directions.}
\end{table}

\subsection{The fluxbrane background}\label{sec:FB}

The motivation for our string construction is to provide a completely geometrical realization of Nekrasov's construction of the equivariant gauge theory~\cite{Nekrasov:2002qd, Nekrasov:2003rj}. For simplicity we will only consider theories on \( \setR^4 \), equivariant with respect to a \( U(1)^2 \) action that we identify with the maximal torus of the \( SO(4) \) symmetry group\footnote{We only consider cases where the $U(1)$ do not act freely.}.

Consider a six-dimensional manifold \( \setR^4 \times T^2 \) with coordinates \( ( \wt x^0, \dots, \wt x^3, \wt x^8, \wt x^9) \). Let \( \wt R_8 \) and \( \wt R_9 \) be the radii of the torus, so that \( \wt x^8 \simeq \wt x^8 + 2 \pi \wt R_8 \) and  \( \wt x^9 \simeq \wt x^9 + 2 \pi \wt R_9 \). The \( U(1)^2 \) action on \( \setR^4 \) is obtained by imposing the following identifications:
\begin{align}
\label{eq:monodromy}
  \begin{cases}
    \wt x^8 \simeq \wt x^8 + 2 \pi \wt R_8 n_8 \\
    \theta_k \simeq  \theta_k + 2 \pi  \epsilon_k^R \wt R_8 n_8
  \end{cases} &&
  \begin{cases}
    \wt x^9 \simeq \wt x^9 + 2 \pi \wt R_9 n_9 \\
    \theta_k \simeq  \theta_k + 2 \pi  \epsilon_k^I \wt R_9 n_9
  \end{cases}
\end{align}
where \( k = 1,2 \), \( n_8, n_9 \in \setZ\), \( \epsilon_k^{R,I} \in \setR \) and  \( \theta_1 = \arctan \wt x^1 / \wt x^0 \), \( \theta_2 = \arctan \wt x^3 / \wt x^2 \) are independently \( 2 \pi \)--periodic variables. In this picture, the six-dimensional, locally flat space is interpreted as an \( \setR^4 \)--fibration over \( T^2 \) with the fibration given by the monodromy in Eq.~\eqref{eq:monodromy}. 

An alternative picture is preferable if one is interested (as we will be in the following) in studying the small-\( \wt R \) limit. In this case one looks at the space as a \( T^2 \)--fibration over \( \setR^4 \). This necessitates the disentanglement of the periodicities of the torus and the angles of the base, which is obtained by introducing new coordinates \( \phi_k \) defined by
\begin{equation}
  \phi_k = \theta_k - \epsilon_k^R \wt x^8 - \epsilon_k^I \wt x^9 = \theta_k - \Re ( \epsilon_k \bar{\wt{v}}),
\end{equation}
where \( \epsilon_k = \epsilon_k^R + \im \epsilon_k^I \) and \( \wt v = \wt x^8 + \im \wt x^9  \).
It is convenient to also introduce a new set of rectangular coordinates given by
\begin{align}
  x^0 + \im x^1 &= \rho_1 \eu^{\im \phi_1}\,, & x^2 + \im x^3 &= \rho_2 \eu^{\im \phi_2} \, .
\end{align}
The metric takes the form
\begin{multline}
\label{eq:fluxbrane-metric}
  \di s^2 = \di \vec x_{0 \dots 3}^2 - \frac{V_i^R V_j^R \di x^i \di x^j}{1 + V^R \cdot V^R} - \frac{V_i^I V_j^I \di x^i \di x^j}{1 + V^I \cdot V^I} \\ 
  + \left( 1 + V^R \cdot V^R \right) \left[ (\di x^8)^2 - \frac{V_i^R \di x^i}{1 + V^R \cdot V^R}  \right]^2 \\ 
  + \left( 1 + V^I \cdot V^I \right) \left[ (\di x^9)^2 - \frac{V_i^I \di x^i}{1 + V^I \cdot V^I}  \right]^2 + 2 V^R \cdot V^I \di x^8 \di x^9 \,,
\end{multline}
where \( V^R \) and \( V^I \) are the generators of the \( U(1) \times U(1) \) rotations in the base, weighted by the \( \epsilon \) parameters:
\begin{align}
  V^{R} &= \epsilon_1^R \left( x^1 \del_0 - x^0 \del_1 \right) +  \epsilon_2^R \left( x^3 \del_2 - x^2 \del_3 \right) ,\\
  V^{I} &= \epsilon_1^I \left( x^1 \del_0 - x^0 \del_1 \right) +  \epsilon_2^I \left( x^3 \del_2 - x^2 \del_3 \right) .
\end{align}
From the expression of the metric we see explicitly the structure of the \( T^2 \)--fibration over \( \setR^4 \) with non-flat metric and connection \( V \).

The construction can be immediately generalized to ten dimensions, where we look at space as a \( T^2 \)--fibration over \( \setR^8 \). This background is referred to as \emph{Melvin} background in general relativity~\cite{Melvin:1963qx} and is the called the \emph{NS fluxbrane} background in the string theory context~\cite{Russo:2001na}. The metric has the same form as in Eq.~\eqref{eq:fluxbrane-metric}, but now the connection takes the form
\begin{multline}
  V = V^R + \im V^I = \epsilon_1 \left( x^1 \del_0 - x^0 \del_1 \right) +  \epsilon_2 \left( x^3 \del_2 - x^2 \del_3 \right) \\
  + \epsilon_3 \left( x^5 \del_4 - x^4 \del_5 \right) +  \epsilon_4 \left( x^7 \del_6 - x^6 \del_7 \right) .
\end{multline}


\subsection{The NS fluxtrap}\label{sec:FT}

Since we are ultimately interested in the study of the four-dimensional gauge theories that describe the dynamics of D--branes in our background, we are interested in the \( \wt R \to 0 \) limit in order to discard the momenta around the torus in the fluxtrap picture.

In a string theory setting this is most easily obtained by T--dualizing the fluxbrane background in \( \wt x^8 \) and \( \wt x^9 \)  and taking the decompactification limit in which the T--dual radii \( R_{8,9} = \alpha'/\wt R_{8,9} \) are very large. The momenta around \( \wt x^8 \) and \( \wt x^9 \) become winding modes around the T--dual directions \( x^8, x^9 \) and decouple in the large--\( R \) limit.

The effect of the two T--dualities is to turn the fluxbrane connection into a \( B \)--field and the metric components \( g_{\wt 8 \wt 8} \) and \( g_{\wt 9 \wt 9} \) into a non-trivial dilaton. This is the so-called \emph{fluxtrap background}:
\begin{align}
  \label{eq:fluxtrap}
  \di s^2 ={}& \di \mathbf{x}^2_{0 \dots 7} + \frac{1}{\Delta^2} \Big[ (\di x^8)^2 + (\di x^9)^2 + \left( V^I \di x^8 - V^R \di x^9 \right)^2 \\
  &- \left( 1 + V^I \cdot V^I \right) V^R V^R + 2 V^R \cdot V^I V^R V^I - \left( 1 + V^R \cdot V^R \right) V^I V^I \Big],\\
  B ={}& \frac{1}{\Delta^2} \left( V^R \wedge \di x^8 + V^I \wedge \di x^9 + V^I \cdot [V^I, V^R ] \wedge \di x^8 + V^R \cdot [V^R, V^I] \wedge \di x^9\right), \\
  \eu^{-\Phi } ={}&  \Delta\,,
\end{align}
where \( \Delta^2 = \left( 1 + V^R \cdot V^R \right) \left( 1 + V^I \cdot V^I \right) - (V^R \cdot V^I)^2 \). 
We see that after the T--duality, the metric is no longer flat, but returns to flat space in the limit of $\epsilon_i\to 0$. The lowest order deformation appears in the B--field, which was generated by the Melvin shifts. The dilaton receives a non-trivial contribution which has a maximum at the origin, thus creating a potential which localizes the instantons at the origin.

\begin{figure}
  \centering
    \begin{tikzpicture}
      \node (0,0) {\includegraphics[scale=.7,trim=0 0 0 18em, clip=true]{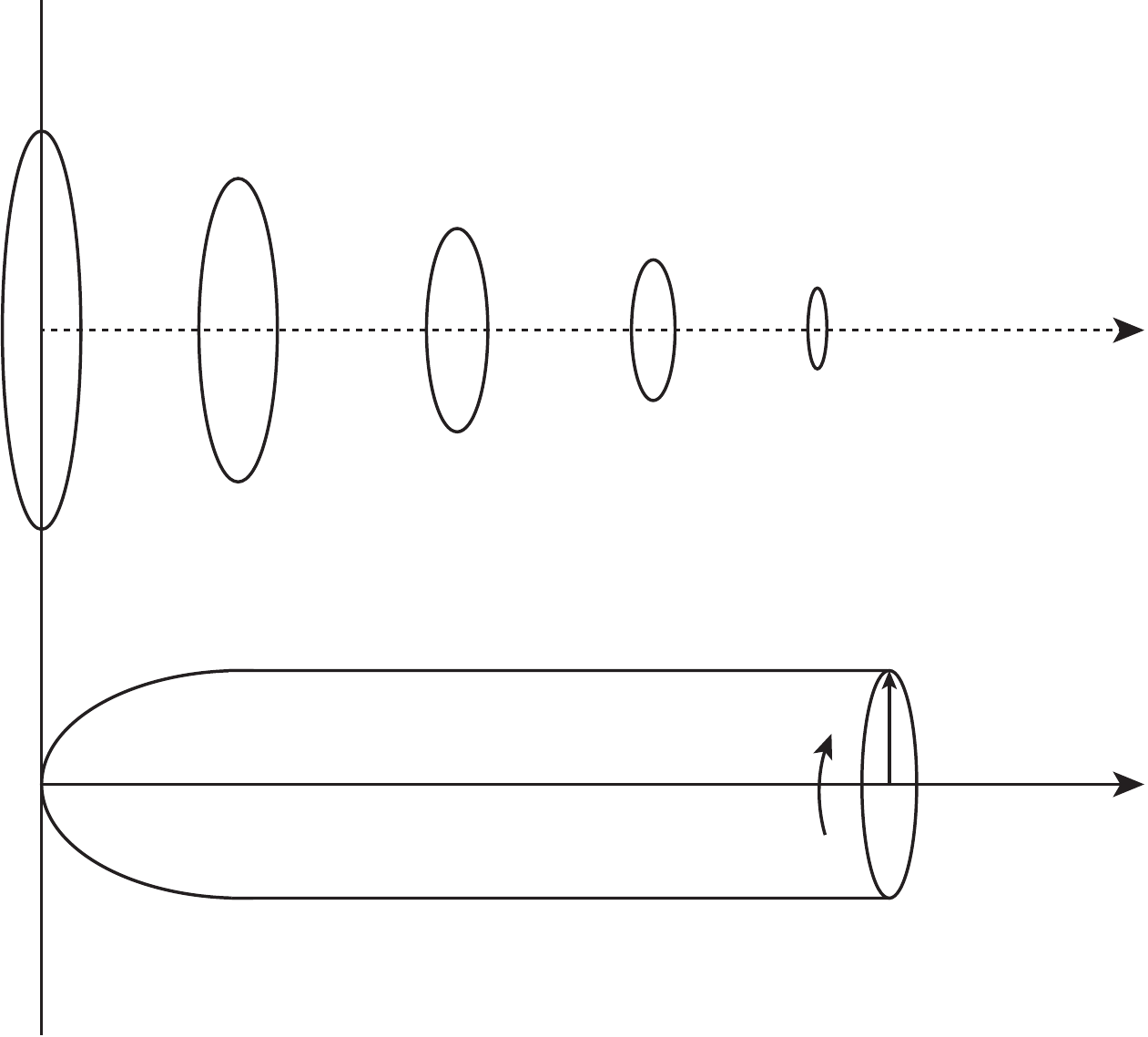}};
      \begin{scope}[shift={(-6,-2)},every node/.style={anchor=base
          west}]
        \begin{small}
        
          \draw (10,1.5) node[]{\( \rho_1 \)}; 
          \draw (8,2.4) node[anchor={base east}]{\( \phi_1  \)};
          \draw (8.6,2.4) node[anchor={base west}]{\( \frac{1}{\epsilon_1} \)};

         

          \draw (1.2,2) node[]{\( \setR^2 \)};

          \draw (8.6,1) node[]{\( \setR \times S^1 \)};
 
        
        \end{small}
      \end{scope}
    \end{tikzpicture}
  \caption{Cartoon of the geometry of the base of the manifold \( M_3( \epsilon_1) \): a cigar with asymptotic radius \( 1/ \epsilon_1 \). }
  \label{fig:cigar-fibration}
\end{figure}
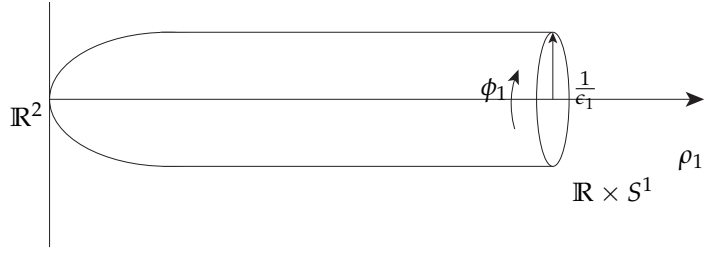

In the special case of \( V^I = 0 \) (\emph{i.e.} for \( \epsilon \) real), the background takes a particularly transparent form~\cite{Hellerman:2011mv, Reffert:2011dp}:
\begin{align}
  \di s^2 &= \delta_{ij} \di x^i\di x^j  + \frac{(\di x^8)^2 - V_i V_j \di x^i \di x^j}{1 + V \cdot V}  \, , \\
  B&=  \frac{V_i\di  x^i \wedge \di x^8}{1 + V \cdot V} \,,\label{eq:BB}\\
  \eu^{-\Phi} &= \sqrt{1 + V\cdot V} \, .
\end{align}
Another interesting situation is obtained if \( V^R \cdot V^I  = 0 \). If for simplicity we set \( \epsilon_1 \in \setR \), \( \epsilon_2 \in \im \setR \), \( \epsilon_3 = \epsilon_4 = 0 \), the background takes the form
\begin{subequations}
  \label{eq:limit-fluxtrap}
  \begin{align}
    \di s^2 &= \di \rho_1^2 + \frac{\rho_1^2 \di \phi_1^2 + \di x_8^2 }{1 + \epsilon_1^2 \rho_1^2} + \di \rho_2^2 + \frac{\rho_2^2 \di \phi_2^2 + \di x_9^2 }{1+ \epsilon_2^2 \rho_2^2} + \sum_{k=4}^7 (\di x^k)^2 ,\\
    B &= \epsilon_1 \frac{\rho_1^2 }{1 + \epsilon_1^2 \rho_1^2} \di \phi_1 \wedge \di x_8 +  \epsilon_2 \frac{\rho_2^2 }{1 + \epsilon_2^2 \rho_2^2} \di \phi_2 \wedge \di x_9 \  , \\
    \eu^{- \Phi} &=  \sqrt{\left( 1 + \epsilon_1^2 \rho_1^2 \right) \left(
        1 + \epsilon_2^2 \rho_2^2 \right)} \  .
  \end{align}
\end{subequations}
The space splits into the product
\begin{equation}
  M_{10} = M_3 (\epsilon_1) \times M_3 (\epsilon_2) \times \setR^4\,,
\end{equation}
where \( M_3 \) is a three-dimensional manifold which is an \( \setR \)--foliation (generated by \( x_8 \) or \( x_9 \)) over the cigar with asymptotic radius \( 1/\epsilon_i \) described by \( (\rho_1, \phi_1) \) or \( (\rho_2, \phi_2) \) (see the cartoon in Figure~\ref{fig:cigar-fibration}):
\begin{equation}
  \begin{tikzpicture}[node distance=5em, auto]
    \node (S) {\( \setR \langle x_8 \rangle \) }; 
    \node (M) [right of=S] {\( M_3 (\epsilon_1)\) };
    \node (cigar) [below of=M] {cigar \(\langle \rho_1, \phi_1 \rangle\)}; 
    \draw[->] (S) to node {} (M); 
    \draw[->] (M) to node {} (cigar);
  \end{tikzpicture}
\end{equation}

This shows that the effect of the \( \Omega \)--deformation is to regularize the rotations generated by \( \del_{\phi_1} \) and \( \del_{\phi_2} \) in the sense that the operators become bounded:
\begin{align}
  \| \del_{\phi_1} \|^2 &= \frac{\rho_1^2}{1 + \epsilon_1^2 \rho_1^2} < \frac{1}{\epsilon_1^2} \  ,&
  \| \del_{\phi_2} \|^2 &= \frac{\rho_2^2}{1 + \epsilon_2^2 \rho_2^2} < \frac{1}{\epsilon_2^2} \  .
\end{align}
In a different frame this will translate into a bound on the asymptotic coupling of the effective gauge theory for the motion of a D--brane.

\subsection{The RR fluxtrap}\label{sec:RRFT}

Up to this point we have considered solutions of bosonic string theory that we are free to embed in either \tIIA or \tIIB. If we choose to look at the fluxtrap as a \tIIB background, we can study its S--dual.
To obtain the S--dual version or \emph{RR fluxtrap} background, we can simply dualize the bulk fields of the fluxtrap background using the standard formulae~\cite{Polchinski:1998rr}:
\begin{equation}
  \begin{aligned}
    \Phi' &= -\Phi \, , & G'_{\mu\nu} &= e^{-\Phi}G_{\mu\nu}\,,\\
    B'_2 &= C_2 \, , & C'_2 &= -B_2.
  \end{aligned}
\end{equation}
We see that while the dilaton goes over to its negative, the S--dual background has no $B$--field, but instead a $C_2$--field which is due to the deformation. The RR 3--form field goes in the leading order in $\epsilon$ with $\omega = \tfrac{1}{2}\, \di V$, which is usually interpreted as a graviphoton field strength. In this sense, the equations above describe the full, non-perturbative backreaction of the fields on the geometry.
This background (or, rather, its small-\( \epsilon \) limit) has been studied in the past in the context of the \( \Omega \)--deformation~\cite{
Hollowood:2003cv, Iqbal:2007ii, Antoniadis:2010iq, Krefl:2010fm,Huang:2010kf, Aganagic:2011mi, Billo:2006jm, Billo:2007va, Billo:2009di}.

We would like to point out that the dynamics of a \D3--brane in the RR--fluxtrap is not governed by the equivariant action of Nekrasov and Okounkov~\cite{Nekrasov:2003rj}, but by its S--dual, which describes a different region of the moduli space.


\subsection{The M--theory fluxtrap}\label{sec:MFT}

When the fluxtrap is seen as a \tIIA background, it can be easily lifted to an eleven-dimensional M--theory background:~\cite{Hellerman:2012zf}
\begin{align}
  G_{IJ} \di x^I \di x^J &= \eu^{-2\Phi/3} g_{ij} \di x^i \di x^j + \eu^{4\Phi/3} \left(\di x^{10} + A_1 \right)^2 \, ,\\
  C_3 &= A_3 + B \wedge \di x^{10} \, .
\end{align}
The lowest order deformation (in \( \epsilon \)) with respect to flat space is in the three-form field \( C_3 \) and is given by
\begin{equation}
  C_3 = \left( V^R \wedge \di x^8 + V^I \wedge \di x^9 \right) \wedge \di x^{10} + \mathcal{O} (\epsilon^3)
\end{equation}
so that  its four-form flux is given by the volume forms over the two-planes weighted by the \( \epsilon_i \):
\begin{equation}
  G_4 = \di C_3 =2 \sum_{k=1}^4 \omega_k \wedge \left( \epsilon_k^R \di x^8 + \epsilon_k^I \di x^9 \right) \wedge \di x^{10} =  \sum_{k=1}^4 \epsilon_k \omega_k \wedge \di \bar v \wedge \di x^{10} + \text{c.c.}\,,
\end{equation}
where \( \omega_k  = \rho_k \di \rho_k \wedge \di \phi_k \) and \( v = x^8 + \im x^9 \).

We will use this background when studying the lift of an \NS5/\D4 system that is needed to derive the effective low energy Lagrangian for the \( \Omega \)--deformed \ac{sym} theory. \textsc{bps} states and surface operators in the gauge theory will be interpreted as \M2--branes and KK modes in this background~\cite{Hellerman:2012rd, Bulycheva:2012ct}. 

\subsection{The M--theory fluxbrane}\label{sec:MFB}

A different M--theory background is obtained when lifting the \tIIA fluxbrane background. In this case the metric is locally flat in the \( \wt x \) coordinates with the identifications in Eq.~(\ref{eq:monodromy}) imposed. Just like the fluxbrane and the fluxtrap are related by T--duality, the M--theory fluxbrane and M--theory fluxtrap are related by the \( S \) element of the \( SL_2(\setZ) \times SL_3(\setZ) \) symmetry group of M--theory compactified on \( T^3 \).
In the case \( V^I = 0 \) one can consider monodromies corresponding to rotations in the direction \( x^{10} \). After writing the locally flat metric in the fluxbrane form of a \( S^1 \)--fibration over \( \setR^9 \times S^1 \) we can reduce on the fiber direction and then T--dualize in the \( S^1 \)--direction of the base. The result is the \tIIB RR fluxtrap of Section~\ref{sec:RRFT}.

The situation is \emph{different} when both \( V^R \) and \( V^I \) are turned on (or, more precisely, when the ratios of the \( \epsilon \)'s are not all real numbers). In this case one can show that there is no frame in which a purely geometrical M--theory description (\emph{i.e.} flat space with identifications without fluxes) is equivalent to the generic M--theory fluxtrap of Section~\ref{sec:MFT} and admits the embedding of an \M5--brane reproducing the \( \Omega \)--deformed \ac{sym} theory that we will obtain in the following (Section~\ref{sec:SW}). For the case of both \( V^R \) and \( V^I \) turned on one \emph{necessarily} has to use the flux\emph{trap} construction, as the fluxbrane cannot generate such a theory.

\subsection{The reciprocal frame}\label{sec:rec}

Starting from the M--theory fluxtrap there are two natural directions in which we can reduce to string theory. Consider for concreteness the case \( V^R \cdot V^I = 0 \) given in Eq.~(\ref{eq:limit-fluxtrap}). While reducing on \( x^{10} \) brings us back to the fluxtrap, we can alternatively reduce on the angles \( \phi_1 \) or \( \phi_2 \). In fact it is interesting to consider the \tIIB background obtained by first reducing on \( \phi_1 \) and then T--dualizing on \( \phi_2 \). This is the \emph{reciprocal frame}~\cite{Hellerman:2012rd}. Here, the bulk fields take the form

\begin{subequations}
  \label{eq:reciprocal-frame}
  \begin{align}
    \begin{split}
      \di s^2 ={} & \epsilon_1 \rho_1 \sqrt{1+\epsilon_2^2\rho_2^2}\left[\di \rho_1^2+\di \rho_2^2+\frac{\di \tilde \sigma_2^2} {\epsilon_1^2 \rho_1^2 \epsilon_2^2 \rho_2^2} + \di \rho_3^2+\rho_3^2\di \psi^2+\di x_6^2+\di x_7^2 +\right.\\
      & \left.+\frac{\di x_8^2}{1+\epsilon_1^2\rho_1^2}+\frac{\di x_9^2}{1+\epsilon_2^2\rho_2^2} +\frac{ \di x_{10}^2}{ \left( 1 +\epsilon_1^2\rho_1^2\right) \left( 1 + \epsilon_2^2\rho_2^2\right)}
      \right] \ ,
    \end{split}
    \\
    B ={} & \frac{ \epsilon_1^2 \rho_1^2}{1+\epsilon_1^2\rho_1^2} \di x_8 \wedge\di x_{10} \ , \\
    \eu^{- \Phi} ={} & \frac{\epsilon_2 \rho_2}{\epsilon_1 \rho_1} \sqrt{\frac{1+ \epsilon_1^2 \rho_1^2}{1 + \epsilon_2^2 \rho_2^2}} \ , \\
    C_2 ={} & \frac{\epsilon_2^2 \rho_2^2}{ 1 + \epsilon_2^2 \rho_2^2} \di x_9 \wedge\di x_{10} \  ,
\end{align}
\end{subequations}
where \( \tilde \sigma_2 \) is periodic with period \( 2 \pi \alpha' \epsilon_2 \).  
The \tIIA background obtained via the reduction in \( \phi_1 \) corresponds to a \D6--brane in the \(\Omega \)--deformed bulk (\emph{i.e.} not a D6 probe brane). The following T--duality in \( \phi_2 \) turns the \D6 into a \D5 bulk brane and also generates an additional \NS5--brane in the background.
The fluxes that appear here are due to the fluxtrap construction and are not  the ones generated by the background branes, which are negligible in the \( \rho_3 \ll \rho_1, \rho_2 \) limit that we are considering. The bulk branes only play the role of boundary conditions for the embedding of the dynamical \D3 probe branes that we will consider in Section~\ref{sec:recg}.
 An interesting feature of this background is that the dilaton vanishes asymptotically for \( \rho_1, \rho_2 \to \infty \).  
This \tIIB background has a simple behavior under S--duality which 
amounts
to exchanging \( \epsilon_1 \) with \( \epsilon_2 \).  This transformation has the effect of swapping the \NS5--brane with the \D5--brane in the bulk.

\bigskip

The intermediate \tIIA background is interesting for a different reason. As we said, it represents a \D6--brane in the \( \Omega \)--deformation which turns on a \( B \)--field. A D--brane with a \( B \)--field admits a dual non-commutative description via the \ac{sw} map. Applying this transformation to the background at hand, we find that the deformed \D6--brane can be equivalently described as a \D6--brane in flat non-commutative space with parameter \( \hbar = \epsilon \)~\cite{Hellerman:2012zf}.

\subsection{Supersymmetry}\label{sec:susy}

One of the advantages of the string theory description of the \( \Omega \)--background is that one can make a simple and direct analysis of the supersymmetry properties in terms of Killing spinors of the ten and eleven-dimensional geometries as opposed to a direct computation of the supersymmetric invariance of the gauge theory~\cite{Hama:2011ea}. The main idea is that one starts with the thirty-two constant Killing spinors of flat space and projects out those that are not compatible with the identifications in Eq.~\eqref{eq:monodromy}.

Let us start with the fluxbrane background, that for concreteness we assume to be embedded in \tIIB. The Killing spinors of flat space can be written in cylindrical coordinates as
\begin{equation}
  \etaB = \left( \mathbf{1} + \Gamma_{11} \right) \prod_{k=1}^N \exp [ \frac{\theta_k}{2} \Gamma_{\rho_k \theta_k} ] \left( \wt \eta_0 + \im \wt \eta_1  \right) \, ,  
\end{equation}
where \( \wt \eta_0 \) and \( \wt \eta_1 \) are constant spinors, \( N \) is the number of planes in which we impose the identifications and \( \Gamma_{\rho_k \theta_k} \) is the product of the gamma matrices in each plane.
The Killing spinor is invariant under \( \theta_k \to \theta_k + 2 \pi n_k \), but not under the Melvin identifications. To isolate the source of the problem we pass to the disentangled coordinates \( \phi_k \),
\begin{equation}
  \etaB = \prod_{k=1}^N \exp [ \frac{\phi_k}{2} \Gamma_{\rho_k \theta_k} ] \exp [\tfrac{1}{2} \Re[\epsilon_k \bar{\wt v}] \Gamma_{\rho_k \theta_k} ] \wt \etaw \, ,
\end{equation}
where $\wt \etaw=\left( \mathbf{1} + \Gamma_{11} \right) \left( \wt \eta_0 + \im \wt \eta_1  \right)$.
For general values of \( \epsilon_k \), the second exponential is not invariant under \( \wt v \mapsto \wt v + 2 \pi n_1 + 2 \pi \im n_2 \), which means that in general \( K  \) is not a good Killing spinor and all supersymmetries are broken. The situation changes when the exponential is singular, \emph{i.e.} when \( N > 1 \) and
\begin{equation}
  \sum_{k = 1}^N \epsilon_k = 0 \, ,
\end{equation}
where the sign of the $\epsilon_k$ reflects the choice of the orientation of the rotation in each of the 2--planes.
Now one can write \( \epsilon_N = - \sum_{k=1}^{N-1} \epsilon_k \) and
\begin{equation}
  \prod_{k=1}^N \exp [\tfrac{1}{2} \Re[\epsilon_k \bar{\wt v}] \Gamma_{\rho_k \theta_k} ] = \prod_{k=1}^{N-1} \exp [\tfrac{1}{2} \Re[\epsilon_k \bar{\wt v}] (\Gamma_{\rho_k \theta_k} - \Gamma_{\rho_N \theta_N} )] \, .
\end{equation}
We have thus obtained the product of \( N - 1 \) commuting matrices, which are annihilated by the projectors
\begin{equation}
  \proj{flux}_k = \tfrac{1}{2} ( \mathbf{1} - \Gamma_{\rho_k \theta_k \rho_N \theta_N}) \, .
\end{equation}
We are now in the position of writing the general expression for a preserved Killing spinor in the fluxbrane background by introducing the spinor \( \etaw \) via
\begin{equation}
  \wt \etaw = \prod_{k=1}^{N-1} \proj{flux}_k \etaw\,,
\end{equation}
so that the following Killing spinor respects the boundary conditions:
\begin{equation}
  \etaB = \prod_{k=1}^N \exp [ \frac{\phi_k}{2} \Gamma_{\rho_k \theta_k} ] \proj{flux}_k \etaw  \, .
\end{equation}
Each projector breaks half of the supersymmetries, thus leaving a total of \( 32/ 2^{N-1} = 2^{6 - N} \) supersymmetries, where \( N \ge 2 \) is the number of deformation parameters.
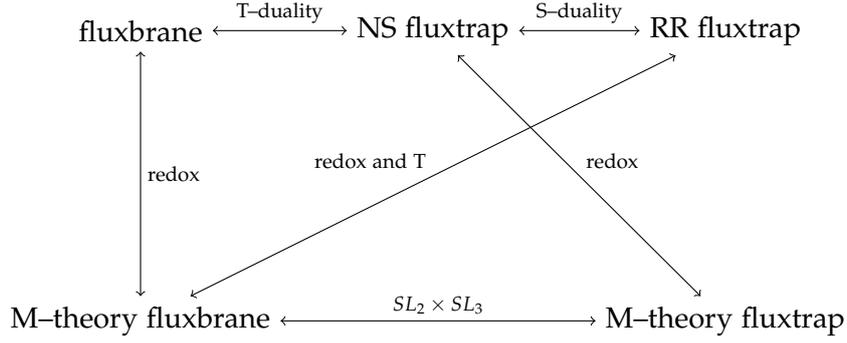
\begin{figure}
  \centering
  \begin{tikzpicture}[node distance=10em and 5em, auto]
    \node (FB) {fluxbrane}; 
    \node (FT) [right of=FB] {NS fluxtrap};
    \node (RRFT) [right of=FT] {RR fluxtrap};
    \node (MFB) [below of=FB] {M--theory fluxbrane}; 
    \node (MFT) [below of=RRFT] {M--theory fluxtrap}; 
    \begin{scope}[every node/.style={scale=.7}]
      \draw[<->] (FB) to node {T--duality} (FT); 
      \draw[<->] (FT) to node {S--duality} (RRFT);
      \draw[<->] (MFB) to node {\( SL_2 \times SL_3 \) } (MFT);
      \draw[<->] (FB) to node {redox} (MFB);
      \draw[<->] (FT) to node {redox} (MFT);
      \draw[<->] (MFB) to node {redox and T} (RRFT);
    \end{scope}
  \end{tikzpicture}
  
  \caption{Relations among the bulk descriptions}
  \label{fig:bulk-dualities}
\end{figure}

Having found the Killing spinors in the fluxbrane, we can translate them into Killing spinors in the other backgrounds by simply following the transformations represented in Figure~\ref{fig:bulk-dualities}.
\begin{itemize}
\item For the fluxtrap, the effect of T--duality is to multiply half of the spinors by the gamma matrices in the directions of the T--dualities:
  \begin{equation}
      \etaB = \prod_{k=1}^N \exp [ \frac{\phi_k}{2} \Gamma_{\rho_k \theta_k} ] \proj{flux}_k \left( \eta_0 + \im \Gamma_8 \Gamma_9 \eta_1\right)  \, ,
  \end{equation}
  where \( \eta_0 \) and \( \eta_1 \) are constant and \( \Gamma_8 \) and \( \Gamma_9 \) are defined as
  \begin{equation}
    \Gamma_8 = g_{8 \mu} e\indices{_a^\mu} \gamma^a\,,
  \end{equation}
  where \( g_{\mu\nu} \) is the fluxbrane metric, \( e\indices{_a^\mu} \) is the inverse vielbein and \( \gamma^a \) are the gamma matrices in flat space.
\item The Killing spinors in M--theory can be obtained by lifting the ones in \tIIA:
  \begin{equation}
    \etaM = \eu^{\Phi/6} \etaA
  \end{equation}
  with an appropriate choice of the eleven-dimensional vielbein~\cite{Hellerman:2012zf}.
\end{itemize}



\section{2d effective field theories with deformations}\label{sec:2d}

As first examples of deformed supersymmetric gauge theories obtained from brane constructions placed into the fluxtrap background, we will consider two-dimensional low energy effective gauge theories with twisted masses. 
One of the uses of this construction is the realization of the two-dimensional gauge/Bethe correspondence~\cite{Nekrasov:2009uh, Nekrasov:2009ui,Chen:2011sj} via string theory~\cite{Orlando:2010uu, Orlando:2010aj, Hellerman:2011mv, Orlando:2011nc}.
\emph{Twisted masses} are terms particular to two-dimensional gauge theories and have their equivalent in the \emph{real masses} of three-dimensional gauge theories. In the superspace formalism, they appear in the Lagrangian as  $\mathscr{L}_{\text{tw}}=\int \di^4\theta \,
  (X^\dagger e^{\theta^-\bar\theta^+\widetilde m_X + \text{h.c.}}X)$,
  where $X$ is a chiral matter field and $e^{\theta^-\bar\theta^+\widetilde m_X}$ are matrices in the
  same representation as $X$ of the maximal torus of the global
  symmetry group.  The twisted mass term cannot
be thought of as a superpotential term, but comes rather
from a deformation of the \textsc{susy} algebra itself.

\subsection{D1--branes: 2d $\Omega$--deformed $\mathcal{N}=(8,8)$ theory with twisted masses }\label{sec:N88}

We start out with a simple, yet extremely general example of a \D1--brane placed into a background with all four $\epsilon$--deformations turned on, see Table~\ref{tab:D1}.
\begin{table}
  \centering
 { \begin{tabular}{lcccccccccc}
    \toprule
    \( x \)   & 0               & 1               & 2               & 3  & 4  & 5  & 6  & 7 & 8  & 9  \\
    \colrule
    fluxtrap & \ep{\epsilon_1} & \ep{\epsilon_2} & \ep{\epsilon_3} &  \ep{\epsilon_4} & \T & \T               \\
    D1--brane & \X &\X & \ep{\phi_1} & \ep{\phi_2} & \ep{\phi_3} & \( \sigma_1 \) & \( \sigma_2 \)  \\
        \botrule
  \end{tabular}
  \label{tab:D1}}
 \tbl{D1--brane and its scalar fields in the fluxtrap background}
\end{table}
With deformations both on and away from the brane worldvolume, the resulting two-dimensional theory is both $\Omega$--deformed and has three twisted mass parameters. Its Lagrangian (from the expansion of the \ac{dbi} action of the \D1--brane to second order in the fields) takes the form
\begin{multline}
  \mathscr{L} = 4 \pi^2 F_{1,2}^2 + \left( \del_\mu \sigma_1 + 2 \pi V^\nu F_{\nu\mu} \right)^2 + \left( \del_\mu \sigma_2 \right)^2 \\ + \sum_{k=1}^3 \left[\left( \delta^{\mu\nu} + V^\mu V^\nu \right) \del_\mu \phi_k \del_\nu\bar \phi_k + \im \epsilon_{k+1} \left( \phi_k V^\mu \del_\mu \bar \phi_k - \text{c.c.}\right) + \epsilon_{k+1}^2 \abs{\phi_k}^2\right]\,,
\end{multline}
where
\begin{equation}
  V = \epsilon_1 \left( x^1 \del_0 - x^0 \del_1 \right) .
\end{equation}
The deformation on the worldvolume gives rise to a covariant derivative with non-minimal coupling for the field $\sigma_1$, an effective metric $g^{ij}=\delta^{ij}+V^{(i} V^{j)}$ for the fields $\phi_1,\,\phi_2\,,\phi_3$ and one-derivative terms which are allowed since Poincar\'e--invariance is broken by the deformation.  The mass terms for the fields $\phi_i$ on the other hand are due to the deformations away from the worldvolume.
The twisted masses break the  $\mathcal{N}=(8,8)$ supersymmetry down to  $\mathcal{N}=(1,1)$. 

Also the case of $\mathcal{N}=(2,2)$ theory with twisted masses discussed in the next section can be understood as a limit of this example where $\epsilon_1 = 0$ and with boundary conditions resulting in  $\phi_2 = \phi_3  = 0$.

\subsection{D2--branes suspended between parallel NS5--branes: 2d $\mathcal{N}=(2,2)$ theory with twisted masses}\label{sec:N22}

2d $\mathcal{N}=(2,2)$ theories with twisted masses play an important role in the gauge/Bethe correspondence~\cite{Nekrasov:2009uh, Nekrasov:2009ui, Orlando:2010uu, Orlando:2010aj}. Their string theory realization was first given in~\cite{Hellerman:2011mv} and extended to brane set-ups reproducing spin chains with $A$-- and $D$--type symmetry groups in~\cite{Orlando:2011nc}.

The simplest case corresponding to symmetry group $su(2)$ is given in Table~\ref{tab:D2} and is realized by a stack of D2--branes suspended between parallel NS5--branes. It is possible to also add $U(L)$--flavor groups to this set-up by adding a stack of $L$ D4--branes~\cite{Orlando:2010uu, Orlando:2011nc}, but we will not discuss this case further and instead direct the reader to the relevant literature.
\begin{table}[h]
 \centering
 { \begin{tabular}{lcccccccccc}
    \toprule
    \( x \)   & 0               & 1               & 2               & 3  & 4  & 5  & 6  & 7 & 8  & 9  \\
    \colrule
    fluxtrap & \ep{\epsilon_1} & \ep{\epsilon_2} & \ep{\epsilon_3} &  & & \T & \T               \\
    D2--brane & \X &\X & \ep{\phi} & & & \X & & \ep{\sigma} \\
    NS5--brane & \X & \X & \X & \X & & & &  & \X& \X \\
        \botrule
  \end{tabular}
  \label{tab:D2}}
   \tbl{D2--brane set-up and its scalar fields in the fluxtrap background}
\end{table}

In the static embedding of the D2--brane, $x^0=\zeta^0,\,x^1=\zeta^1,\,x^6=\zeta^3$, the equations of motion are solved for the D--branes sitting in $x^2=x^3=x^4=x^5=x^7=0$. The fluxtrap geometry thus \emph{traps} the D--branes at the origin. In the following, we will briefly discuss the simple case where $\epsilon_1=0,\, \epsilon_2=-\epsilon_3=m$, where $m$ is a real parameter. Expanding the \ac{dbi} action to second order in the fields, we arrive at the following low energy effective Lagrangian for the gauge theory~\cite{Hellerman:2011mv}:
\begin{equation}
 \mathscr{L}_{m}  = 
 - \frac{1}{4\,g^2}\int \mathrm{d}^3 \zeta \left[\del_\mu\sigma\del_\mu\bar\sigma+\del_\mu\phi\del_\mu\bar\phi + m^2  |\phi|^2  + \mathrm{fermions}\right].
\end{equation}
The $\epsilon$--deformation of the string theory bulk is thus inherited by the gauge theory as a mass deformation for the scalar field $\phi$ which encodes the fluctuations of the D2--brane in the $2$ and $3$ directions.

The background deformed by $\epsilon_1=0,\, \epsilon_2=-\epsilon_3=m$ preserves 16 supercharges. Adding the D2--branes and the NS5--branes breaks each another half of the supercharges. We are thus left with four real supercharges, resulting in $\mathcal{N}=(2,2)$ supersymmetry in the effective gauge theory.

\section{4d effective field theories with deformations}\label{sec:4d}

Brane configurations leading to a four-dimensional effective field theory on the brane world-volume are either stacks of D3--branes in type IIB string theory or D4--branes suspended between NS5--branes in type IIA string theory.  $\mathcal{N}=2$  \ac{sym} is the theory where the $\Omega$--deformation was first introduced~\cite{Nekrasov:2002qd, Nekrasov:2003rj}. The special case of $\epsilon_1=-\epsilon_2$ reproduces the topological string partition function, whereas the case $\epsilon_2=0,\, \epsilon_1=-\epsilon_3=m$ corresponds to the Nekrasov--Shatashvili limit of the 4d gauge/Bethe correspondence~\cite{Nekrasov:2009rc}.

Deformations of $\mathcal{N}=4$  \ac{sym} theory can be realized via the fluxtrap construction as well and can serve as a starting point for the construction of their gravity duals via the AdS/CFT correspondence.

Lastly, even $\Omega$--deformed $\mathcal{N}=1$ theory can be realized from the fluxtrap background via a modified brane set-up.

\subsection{D3--branes: 4d $\Omega$--deformed $\mathcal{N}=4$  \ac{sym} and $\mathcal{N}=2^*$}\label{sec:N4}

The four-dimensional $\Omega$--deformed $\mathcal{N}=4$ \ac{sym} theory was described in~\cite{Hellerman:2012rd}.\footnote{An alternative, inequivalent \( \Omega \)--deformation of the same theory was presented in~\cite{Ito:2011wv, Ito:2012hs, Ito:2013eva}.} We consider the brane configuration given in Table~\ref{tab:D3}.
\begin{table}
\centering
 { \begin{tabular}{lcccccccccc}
    \toprule
    \( x \)   & 0               & 1               & 2               & 3  & 4  & 5  & 6  & 7 & 8  & 9  \\
    \colrule
    fluxtrap & \ep{\epsilon_1} & \ep{\epsilon_2} & \ep{\epsilon_3} &  \ep{\epsilon_4} & \T & \T               \\
    D3--brane & \X &\X & \X &\X  & \ep{\phi_1} & \ep{\phi_2} & \ep{\phi_3} \\
        \botrule
  \end{tabular}
  \label{tab:D3}}
    \tbl{D3--brane in the fluxtrap background with scalar fields}
\end{table}
The Lagrangian obtained from the expansion of the \ac{dbi} action is given by
\begin{multline}\label{eq:neq4}
  \mathscr{L}_{\Omega} = \frac{1}{4\gYM^2}\Bigg[ F_{ij}F^{ij} + 
 \frac{1}{4} \left( \delta^{ij} + V^i \bar V^j \right) \left( \del_i \phi_1 \del_j \bar \phi_1 + \text{c.c.} \right) + \frac{1}{4} \left( \delta^{ij} + V^i \bar V^j \right) \left( \del_i \phi_2 \del_j \bar \phi_2 + \text{c.c.} \right)  \\
 + \frac{1}{2} \left( \del^i \phi_3 + V^k F\indices{_{k}^{i}}\right) \left( \del_i \bar \phi_3 + \bar V^kF_{ki} \right)  - \frac{1}{8}{( \bar V^i \del_i \phi_3 - V^i \del_i \bar \phi_3 + V^k\bar V^l F_{kl} )}^2 \\
 + \frac{1}{2 \im}  \left( \epsilon_3 \bar V^i + \bar \epsilon_3 V^i \right)  \left( \bar \phi_1 \del_i  \phi_1 - \text{c.c.} \right) + \frac{1}{2} \abs{\epsilon_3}^2 \phi_1 \bar \phi_1 \\
 +  \frac{1}{2 \im}  \left( \epsilon_4 \bar V^i + \bar \epsilon_4 V^i \right)  \left( \bar \phi_2 \del_i  \phi_2 - \text{c.c.} \right) + \frac{1}{2} \abs{\epsilon_4}^2 \phi_2 \bar \phi_2 \Bigg] \  ,
\end{multline}
where \( V  = \epsilon_1 \left( \xi^0 \del_1 - \xi^1 \del_0 \right) + \im \epsilon_2 \left( \xi^2 \del_3 - \xi^3 \del_2 \right) \) and \( \gYM^2 = 2 \pi \gBO \). 
The deformation results in an effective metric $g^{ij}=\delta^{ij}+V^{(i}\bar V^{j)}$ for the fields $\phi_1,\,\phi_2$. Moreover, these fields acquire mass terms and a one-derivative term, which is allowed by the broken Poincaré invariance. 

$\mathcal{N}=2^*$ theory is a limit of the above general case, namely the case of $\epsilon_1=\epsilon_2=0$, which results in $V=0$, and $\epsilon_3=\epsilon_4=\epsilon$. The $\mathcal{N}=4$ Lagrangian of Eq.~(\ref{eq:neq4}) reduces to
\begin{multline}
  \mathscr{L}_{\Omega} = \frac{1}{4\gYM^2}\Bigg[ F_{ij}F^{ij} 
 + \frac{1}{2} \sum_{k=1}^3 \del^i \phi_k  \del_i \bar \phi_k   + \frac{1}{2} \abs{\epsilon}^2 \phi_1 \bar \phi_1 + \frac{1}{2} \abs{\epsilon}^2 \phi_2 \bar \phi_2 \Bigg] \, .
\end{multline}
We see that the scalar fields $\phi_1$ and $\phi_2$ have received mass terms from the $\epsilon$--deformation, while $\phi_3$ has remained massless.

\subsection{D4--branes suspended between parallel NS5--branes: 4d $\Omega$--deformed $\mathcal{N}=2$  SYM}\label{sec:N2}

\begin{table}[h]
  \centering
{  \begin{tabular}{lcccccccccc}
    \toprule
    \( x \)   & 0               & 1               & 2               & 3  & 4  & 5  & 6  & 7 & 8  & 9  \\
    \colrule
    fluxtrap & \ep{\epsilon_1} & \ep{\epsilon_2} & \ep{\epsilon_3} &  & & \T & \T               \\
    D4--brane & \X &\X  &\X &\X  & & & \X & & \ep{\phi} \\
    NS5--brane & \X & \X & \X & \X & & & &  & \X& \X \\
        \botrule
  \end{tabular}
  \label{tab:D4}}
   \tbl{D4--brane set-up in the fluxtrap background and its scalar field}
\end{table}
Suspending D4--branes between parallel NS5--branes gives instead rise to $\Omega$--deformed $\mathcal{N}=2$ \ac{sym} theory, see Table~\ref{tab:D4}.
Also this case can be obtained as a limit of the 4d $\Omega$--deformed $\mathcal{N}=4$ Lagrangian~(\ref{eq:neq4}). Since the D4--brane is fixed to the NS5--brane in the $4,\,5,\,6,\, 7$ directions, the fluctuations in these directions are zero, $\phi_1=\phi_2=0$. The Lagrangian thus results in 
\begin{multline}\label{eq:neq2}
  \mathscr{L}_{\Omega} = \frac{1}{4\gYM^2}\Bigg[ F_{ij}F^{ij} + 
 \frac{1}{2} \left( \del^i \phi + V^k F\indices{_{k}^{i}}\right) \left( \del_i \bar \phi + \bar V^j F_{ji} \right)\\
   - \frac{1}{8}{( \bar V^i \del_i \phi - V^i \del_i \bar \phi + V^k\bar V^l F_{kl} )}^2  \Bigg] \,.
\end{multline}
The Lagrangian given above is a slightly more general case than the one first given in Eq.~(3.10) of~\cite{Hellerman:2012zf}.

\subsection{D4--branes suspended between non-parallel NS5--branes: 4d $\Omega$--deformed $\mathcal{N}=1$ theory}\label{sec:N1}

While the previous examples were all based on a similar brane placement, realizing $\Omega$--deformed $\mathcal{N}=1$ theory in four dimensions requires a different set-up with \D4--branes suspended between \NS5--branes that are not parallel. This in turn adds a new constraint on the choice of the (dual) Melvin directions which should be parallel to both \NS5--branes. The only possible configuration is the one in Table~\ref{tab:D42}, where only three \( \epsilon \)'s are possible, since there is no \( U(1) \) symmetry in the \( (x^6, x^7) \)--plane. Note that in this example the dynamical \D4--brane is extended in the dual Melvin directions, thus further breaking Lorentz invariance.
The system preserves two real supercharges.

The \ac{dbi} action for the \D3--brane provides the \( \Omega \)--deformation of \( \mathcal{N} = 1 \) \ac{sym}:
\begin{equation}
  \mathscr{L}_{\Omega} = \frac{1}{4 g^2} F_{ij} F^{ij} + V^R_i F^{ij} \mathbf{e}^8_j + V^I_i F^{ij} \mathbf{e}^9_j \,,
\end{equation}
where \( \mathbf{e}^8 \) and \( \mathbf{e}^9 \) are the unit vectors in the directions \( x^8 \) and \( x^9 \), \emph{i.e.} \( \mathbf{e}^8 = \mathbf{e}^8_i \di x^i = \di x^8 \).

\begin{table}
 \centering
{  \begin{tabular}{lcccccccccc}
    \toprule
    \( x \)   & 0               & 1               & 2               & 3  & 4  & 5  & 6  & 7 & 8  & 9  \\
    \colrule
    fluxtrap & \ep{\epsilon_1} & \ep{\epsilon_2} & \ep{\epsilon_3} &  && \T & \T               \\
    D4--brane & \X &\X  & &  & & & \X & &\X &\X \\
    NS5--brane 1 & \X & \X & \X & \X & & & &  & \X& \X \\
    NS5--brane 2 & \X & \X & & & \X & \X & &  & \X& \X \\
        \botrule
  \end{tabular}
  \label{tab:D42}}
  \tbl{D4--brane set-up in the fluxtrap background, no scalar fields are present}
\end{table}

\subsection{$\Omega$--deformed Seiberg--Witten Lagrangian}\label{sec:SW}

The \ac{sw} action can be obtained as the effective four-dimensional action for the flat space embedding of a \M5--brane on a Riemann surface~\cite{Lambert:1997dm}. Repeating the same construction in the M--theory fluxtrap background that we have described in Section~\ref{sec:MFT} leads to the \( \Omega \)--deformation of the \ac{sw} action, \emph{i.e.} the effective low energy action for the \( \Omega \)--deformation of \( \mathcal{N} = 2 \) \ac{sym}~\cite{Lambert:2013lxa}.

The idea is as follows. Start from the supersymmetric embedding of the \M5--brane (\( \setR^4 \times \Sigma \)) and deform it in an appropriate way. The six-dimensional equations of motion (requiring that the \M5 is a generalized minimal surface and that the self-dual three-form field is the pullback of the bulk field) constrain the dynamics of the fluctuations. Integrating the equations over the Riemann surface leads to four-dimensional space-time equations. Finally, these are interpreted as the extremization of an action, \emph{i.e.} as Euler--Lagrange equations.


If we consider the leading-order deformation, the supersymmetric \M5--brane embedding in the fluxtrap is still of the type \( \setR^4  \times \Sigma \)~\cite{Hellerman:2012zf}. 
%
Now we have to deform this embedding: since we are interested in the effective four-dimensional theory living on \( x^0, \dots, x^3 \) which results from integrating the \M5 equations of motion over the Riemann surface \( \Sigma \), we will assume that:
\begin{enumerate}
\item the geometry of the \M5--brane is still a fibration of a Riemann surface over \( \setR^4 \);
\item for each point in \( \setR^4 \) we have the same Riemann surface as above, but with a different value of the modulus \( u \).
\end{enumerate}
In other words, the modulus \( u \) of \( \Sigma \) is a function of the worldvolume coordinates and the embedding is still formally defined by the same equation, but now \( s = s ( z | u(x^\mu)) \) so that the \( x^\mu \)--dependence is entirely captured by
\begin{equation}
  \label{eq:xmu-dependence}
  \del_\mu s (z | u (x^\mu)) = \del_\mu u \frac{\del s}{\del u} \,.
\end{equation}

We ultimately want to discuss the gauge theory living on the worldvolume coordinates \( x^0, \dots, x^3 \). We therefore make the following self-dual ($\im*_6 \Phi =   \Phi$) ansatz for the field \( \Phi \) describing the fluctuations of the three-form living on the brane:
\begin{equation}
  \begin{aligned}
    \Phi ={}&\frac{ \kappa}{2}\cF_{\mu\nu}\di x^\mu\wedge \di x^\nu\wedge\di z+ \frac{\bar \kappa}{2}\wt{\cF}_{\mu\nu}\di x^\mu\wedge \di x^\nu\wedge\di \bar z\\
    & +  \frac{1}{1+|\del s|^2}\frac{1}{3!}\epsilon_{\mu\nu\rho\sigma}\left(  \del^\tau s \bdel \bar s \, \kappa \cF_{\sigma\tau}-\del^\tau\bar s\del s \, \bar\kappa \wt{\cF}_{\sigma\tau} \right)\di x^\mu\wedge\di x^\nu\wedge\di x^\rho \,.
  \end{aligned}
\end{equation}
The two-form \( \cF \) is anti-self-dual in four dimensions, while \( \wt{\cF} \) is self-dual.
\( \kappa(z) \) is a holomorphic function given by~\cite{Lambert:1997dm}
\begin{equation}
  \kappa = \frac{\di s}{\di a} =
  \left( \frac{\di a}{\di u}  \right)^{-1} \lambda_z \, .
\end{equation}
Here $\lambda = \lambda_z dz$ is the holomorphic one-form on $\Sigma$ and \( a \) is the scalar field used in the \ac{sw} solution and related to $\lambda$ by
\begin{equation}
  \frac{\di a}{\di u} = \oint_A \lambda\ ,
\end{equation}
where $A$ is the a-cycle of $\Sigma$. In the following,  ${\cF}$ and $\wt{\cF}$ will be related to the four-dimensional gauge field strength, thus justifying our ansatz.

The \emph{vector equation} is obtained by requiring the differential of the three-form \( h \) that lives on the \M5--brane to be the pullback of the four-form flux in the bulk. Concretely we write \( h = -\tfrac{1}{4} (\hat C_3 + \im * \hat C_3 + \Phi ) \) and impose the condition \( d h = - \tfrac{1}{4} \hat H_4 \).  This becomes 
\begin{equation}
  \di \Phi = \im \di * \hat C_3\,,
\end{equation}
showing that the bulk form acts as a source for the fluctuations.
To obtain the equations of motion of the vector zero-modes in four dimensions we need to reduce these equations on the Riemann surface, which is possible because they can be written as the vanishing of a holomorphic and an anti-holomorphic one-form on \( \Sigma \). 
The final form of the four-dimensional vector equations is
\begin{multline}
  \label{eq:vector-equation}
  \left( \tau - \bar \tau \right) \left[ \del_\mu F_{\mu\nu} + \tfrac{1}{2} \del_\mu ( a + \bar a ) \hat\omega_{\mu\nu} + \tfrac{1}{2} \del_\mu (a - \bar a) \dual{\hat\omega}_{\mu\nu} \right] \\+ \del_\mu \left( \tau - \bar \tau\right) \left[ F_{\mu\nu} + \tfrac{1}{2} \left( a - \bar a \right) \dual{\hat\omega}_{\mu\nu} \right]  
  - \del_\mu
  \left( \tau + \bar \tau \right) \left[ \dual{F}_{\mu\nu} + \tfrac{1}{2} \left( a - \bar a \right)  \hat\omega_{\mu\nu} \right] = 0 \, ,
\end{multline}
where \( \dual{F} = *_4 F \) is related to \( \cF \) by the condition \( \cF = (1 - *) F - (a - \bar a) \omega^-\).  

The covariant equations of motion for the \M5--brane~\cite{Howe:1996yn,Howe:1997fb} in linear order in $\epsilon$ and quadratic order in spatial derivatives $\del_\mu$ is
\begin{equation}
\label{eq:general-scalar-equation}
 \left( \hat g^{mn} - 16 h^{mpq}h^n{}_{pq} \right) \nabla_m \nabla_n X^{I} =  -\frac{2}{3 }  \hat G\indices{^{I}_{mnp}} h^{mnp} \,,
\end{equation}
where $I = 6,\dots,10$ and the geometrical quantities are defined with respect to  the pullback of the spacetime metric to the brane ${\hat g}_{mn}$. This can be interpreted as the vanishing of two scalar densities (corresponding to the fluctuations in \( z  \) and \( \bar z \)) on \( \Sigma \) which can be integrated using \( \lambda \). After the integration, the scalar equations take the final form
\begin{align}
  \begin{multlined}[.9\textwidth]
    \label{eq:scalar-equation-a}
    \left( \tau - \bar \tau \right) \del_\mu \del_\mu a + \del_\mu a  \del_\mu \tau + 2 \frac{\di \bar \tau}{\di \bar a} \left( F_{\mu\nu} F_{\mu\nu} + F_{\mu\nu} \dual{F}_{\mu\nu} \right) \\
    + 4 \frac{\di \bar \tau}{\di \bar a} \left( a - \bar a \right) \hat\omega^+_{\mu\nu} F_{\mu \nu} - 4 \left( \tau - \bar \tau \right) \hat\omega^-_{\mu\nu} F_{\mu\nu} = 0\; ,
  \end{multlined}\\
  \begin{multlined}[.9\textwidth]
    \label{eq:scalar-equation-abar}
    \left( \tau - \bar \tau \right) \del_\mu \del_\mu \bar a - \del_\mu \bar a \del_\mu \bar \tau - 2 \frac{\di \tau}{\di a} \left( F_{\mu\nu} F_{\mu\nu} - F_{\mu\nu} \dual{F}_{\mu\nu} \right) \\
    + 4 \frac{\di \tau}{\di a} \left( a - \bar a \right) \hat\omega^-_{\mu\nu} F_{\mu \nu} - 4 \left( \tau - \bar \tau \right) \hat\omega^+_{\mu\nu} F_{\mu\nu} = 0\; .
  \end{multlined}
\end{align}
These consistent results justify our previous ansatz and assumptions. The four-dimensional vector equation Eq.~(\ref{eq:vector-equation}) and scalar equations Eq.~(\ref{eq:scalar-equation-a}) and (\ref{eq:scalar-equation-abar}) turn out to be Euler--Lagrange equations for a four-dimensional action.

The generalization to arbitrary gauge group and matter content is given by
\begin{multline}
 \im  \mathscr{L} = - \left( \tau_{ij} - \bar \tau_{ij} \right) \Big[ \tfrac{1}{2}\left( \del_\mu a^i + 2\left(\tfrac{  \bar \tau }{\tau - \bar \tau}\right)_{ik} \dual{F}^{k}_{\mu\nu} \dual{\hat U}_\nu \right) \left( \del_\mu \bar a^j -2 \left(\tfrac{  \tau }{\tau - \bar \tau}\right)_{jl} \dual{F}^{l}_{\mu\nu} \dual{\hat U}_\nu \right) \\
  + \left( F^i_{\mu\nu} + \tfrac{1}{2} \left( a^i - \bar a^i \right) \dual{\hat\omega}_{\mu\nu} \right) \left( F^j_{\mu\nu} + \tfrac{1}{2} \left( a^j - \bar a^j \right) \dual{\hat\omega}_{\mu\nu} \right) \Big] \\
+ \left( \tau_{ij} + \bar \tau_{ij} \right) \left( F^i_{\mu\nu} + \tfrac{1}{2} \left( a^i - \bar a^i \right) \dual{\hat\omega}_{\mu\nu} \right) \left( \dual{F}^{j}_{\mu\nu} + \tfrac{1}{2} \left( a^j - \bar a^j \right) \hat\omega_{\mu\nu}  \right)\ ,
\end{multline}
where we have used a suitable form for the inverse of $(\tau - \bar\tau)_{ij}$ which is taken to act from the left. We see that the fluxtrap deformation has generated a generalized covariant derivative for the scalar a with non-minimal coupling to the gauge field and a shift in the gauge field strength for the vector field. The above result does not depend on the compactification radius to type IIA string theory, which is
related to the gauge coupling in four dimensions. It captures therefore all orders in gauge theory and is a quantum result, despite being purely classical from the M--theory point of view. It moreover applies to any Riemann surface. In the case of general Riemann surfaces, there exists an alternative inequivalent orientation of the background field leading to the effective S--dual theory.

\subsection{The reciprocal gauge theory}
\label{sec:recg}

The usual interpretation of the \ac{agt} correspondence is that it relates a \( \Omega \)--deformed gauge theory on \( S^4 \) to a Liouville field theory on a Riemann surface \( \Sigma \) because the two theories can be understood as the reductions to four and two dimensions of an \M5--brane wrapped on \( S^4_{\epsilon_1, \epsilon2} \times \Sigma \). We are not yet able to reproduce such a theory, but the construction in the previous section realizes a close relative corresponding to an \M5--brane wrapped on \( \setR^4_{\epsilon_1,\epsilon_2} \times T^2 \). It is interesting to study the reduction of this configuration on the two angular isometries of \( \setR^4 \) to go to the so-called reciprocal frame
~\cite{Hellerman:2012rd}, see Table~\ref{tab:Everything}.
The corresponding reciprocal gauge theory forms a good starting point for a string theory realization of the \ac{agt} correspondence~\cite{Alday:2009aq}, as it reproduces certain key characteristics of Liouville theory: its loop-counting parameter is \( b^2 = \epsilon_2 / \epsilon_1 \) and S--duality is realized as the exchange \( b \leftrightarrow 1/b \). 
\begin{table}
\centering
 { \begin{tabular}{llccccccccccc}
    \toprule
    frame                                                                   & object   & $x_0$  & $x_1$       & $x_2$  & $  x_3$         & $x_4$ & $x_5$ & $x_6$      & $x_7$     & $x_8$     & $x_9$     & $x_{10}$       \\  \colrule
{M--theory}      & \M5      & $\times$  & $\times$       & $\times$  & $\times$           &          &        & $\times$   &           &           &           & $\times$       \\      \colrule
 {reciprocal frame}                                    & \D3      & $\times $ & $\blacksquare$ & $\times $ &                    &          &        & $\times $  &           &           &           & $\times$       \\ 
                                     
                                      & \D5      &           & $\blacksquare$ & $\times $ &                    &          &        & $\times $  & $\times $ & $\times $ & $\times $ & $\times$       \\ 
     & \NS5     & $\times $ & $\blacksquare$ &           &                    &          &        & $\times $  & $\times $ & $\times $ & $\times $ & $\times$       \\ 
    \botrule
  \end{tabular}
  \label{tab:Everything}}
 \tbl{Extended objects in different frames. The direction marked with a square (\( \blacksquare \)) in type~\textsc{ii} is not geometrical. The D5-- and NS5--branes in the reciprocal frame are non-dynamical objects in the bulk which only appear as a consequence of the reduction from M--theory and duality along angular directions.}
\end{table}
The reduction to \tIIA turns the \M5 into a \D4--brane and the T--duality finally leads to a \D3--brane in the reciprocal background (see Table~\ref{tab:Everything}). The effective theory of this brane is what we call the reciprocal gauge theory.

Consider the static embedding for the D--brane extended in \( \rho_1, \rho_2, x_6, x_{10} \):
\begin{align}
  \rho_1 &= y_1 \ , & \rho_2 &= y_2 \ , & x_6 &= y_3 \ , & x_{10} &= y_4 \ .
\end{align}
The geometry seen by the \D3--brane is that of a two-torus fibration (generated by \( y_3, y_4 \)) over \( \setR^2_+ \)  (generated by \( y_1, y_2 \)):
\begin{equation}
  \begin{tikzpicture}[node distance=5em, auto]
    \node (T2) {\( T^2 \langle y_3, y_4 \rangle\) }; 
    \node (M) [right of=S] {\( M_4 \) };
    \node (R2) [below of=M] { \( \setR^2_+ \langle y_1, y_2 \rangle \)}; 
    \draw[->] (T2) to node {} (M); 
    \draw[->] (M) to node {} (R2);
  \end{tikzpicture}
\end{equation}
\def\LOCVAR{U}
The dynamics is described by the fields
\begin{align}
  \LOCVAR_1 + \im \LOCVAR_2 &= \frac{\rho_3 \eu^{\im \psi}}{2 \pi \alpha'} \ , & \LOCVAR_3 &= \frac{x_7}{2 \pi \alpha'} \ , & \LOCVAR_4 &= \frac{\tilde \sigma_2 }{2 \pi \alpha'} \ , & \LOCVAR_5 &= \frac{x_8}{2 \pi \alpha'} \ , & \LOCVAR_6 &= \frac{x_9}{2 \pi \alpha'} \  .
\end{align}
The effective action for the \D3--brane is given by
\begin{multline}
  \mathscr{L}_{\text{rec}} = \frac{y_2}{8 \pi y_1} F_{kl} F_{kl} + \frac{\epsilon_1^2 y_1 y_2}{4 \pi} \Bigg[ \sum_{k=1}^3 {\left( F_{k4} - \del_k \LOCVAR_5 \right)}^2 + \frac{1}{\Delta_2^2} \sum_{k=1}^3 {\left( \im \frac{\epsilon_2}{\epsilon_1} \frac{y_2}{y_1} {(*F)}_{k4} - \del_k \LOCVAR_6 \right)}^2 \\ 
+ \tau^{kl} (\xi) h^{ij} (\xi) \del_k \LOCVAR_i \del_l \LOCVAR_j  +  \Delta_2^2 {(\del_4 \LOCVAR_5)}^2 + \Delta_1^2 {( \del_4 \LOCVAR_6)}^2 + \left( y_1^{-2} + y_2^{-2} \right) \left( \LOCVAR_1^2 + \LOCVAR_2^2 \right) \Bigg],
\end{multline}
where
\begin{align}
  \tau^{kl}(\xi) &=
  \begin{pmatrix}
    1 \\ & 1 \\ && 1 \\ &&& \Delta_1^2 \Delta_2^2
  \end{pmatrix} \  ,&
 h^{ij}(\xi) &=
  \begin{pmatrix}
    1 \\ & 1 \\ && 1 \\ &&& {\left(\epsilon_1 y_1\right)}^{-2} {\left(\epsilon_2 y_2\right)}^{-2}
  \end{pmatrix}.
\end{align}
In order to study the effective gauge coupling, it is convenient to define the gauge kinetic \emph{tensor} \( M^{ijkl} \) from
\begin{equation}
  \mathscr{L}_g =  M^{ijkl} F_{ij} F_{kl} \  ,
\end{equation}
and the scalar \( \geff =  \sqrt{ \frac{2}{3} \epsilon_{ijkl} \epsilon_{i'j'k'l'} M^{iji'j'} M^{klk'l'} } \).
In our case, we find that the effective gauge coupling of the reciprocal theory takes the form
\begin{equation}
  \frac{1}{\grec^2} = \frac{1}{2 \pi} \frac{y_2 \sqrt{1 + \epsilon_1^2 y_1^2}}{y_1 \sqrt{1 + \epsilon_2^2 y_2^2} } \xrightarrow[y_1, y_2 \to \infty]{} \frac{\epsilon_1}{2 \pi \epsilon_2} .
\end{equation}
We see thus that the asymptotic gauge coupling is given by the ratio of the two $\epsilon$--parameters as it is the case in the Liouville theory in the \ac{agt} correspondence.

In order to study the behavior of the action under S--duality we need a notion of inverse coupling. Then we can define the S--dual as the action obtained by inverting the 
tensor\footnote{See~\cite{Hellerman:2012rd} for a suitable definition of the inverse.} \( M \) and dualizing the gauge field:
\begin{equation}
  \mathscr{L}_{\text{dual}} = \frac{1}{16 \pi^2} ( M^{-1} )^{ijkl} (*F)_{ij} (*F)_{kl} \  .  
\end{equation}
In our case  \( M \) is a symmetric matrix and the action has been written explicitly in terms of the gauge field and its dual. It follows that
\begin{equation}
  \mathscr{L}_{\text{dual}}(\epsilon_1, \epsilon_2) = \frac{y_1}{4 \pi y_2} \left[ \frac{{(*F)}_{4k} {(*F)}_{k4}}{1 + \epsilon_1^2 y_1^2 } + F_{k4} F_{k4}\left( 1 + \epsilon_2^2 y_2^2 \right) \right]  .
\end{equation}
It is immediate to see that the effect of S--duality is simply to exchange \( \epsilon_1 \) and \( \epsilon_2 \) as we had already observed at the string level by looking at the reciprocal frame:
\begin{equation}
  \boxed{\mathscr{L}_{\text{dual}} (\epsilon_1, \epsilon_2) = \mathscr{L}_g(\epsilon_2, \epsilon_1) \  .  }
\end{equation}

In the \ac{agt} correspondence one identifies the Liouville parameter \( b \) with the ratio of the two epsilons,
\begin{equation}
  b^2 = \frac{\epsilon_2}{\epsilon_1} \ .  
\end{equation}
Even though the reciprocal gauge theory is intrinsically four-dimensional, we have thus seen that it shares at least two remarkable properties with the two-dimensional Liouville field theory:
\begin{enumerate}
\item The asymptotic coupling constant is proportional to \( b^2 \);
\item S--duality exchanges \( b \leftrightarrow 1/b \),  just like the Liouville duality that exchanges the perturbative and the instanton spectrum.
\end{enumerate}


\subsection{The AdS/CFT dual}\label{sec:AdSCFT}

Since we have string realizations of deformations of \( \mathcal{N} = 4 \) \ac{sym} based on the dynamics of a \D3--brane, it is natural to look for a construction of the gravity dual of the \( \Omega \)--deformed theory. We have seen in particular that as a special case, the fluxtrap provides a construction for \( \mathcal{N} = 2^* \) theory. Gravity duals of massive deformations have already been studied extensively in the literature, starting from the work of Polchinski and Strassler~\cite{Polchinski:2000uf}. In fact the lowest order deformation of the \D3--background found in~\cite{Polchinski:2000uf} is given by a three-form flux that coincides precisely with the one in the fluxtrap of Eq.~(\ref{eq:fluxtrap}). We conclude that the gravity dual of the \( \Omega \)--deformed \ac{sym} is given by the full backreaction of the \D3--brane in the fluxtrap, which interpolates between the solution of Polchinski and Strassler in the near-horizon limit and the flat-space fluxtrap of Eq.~(\ref{eq:fluxtrap}) at infinity.

We have evaluated the solution at first order in \( \epsilon \) and part of the second order in two special cases:
\begin{enumerate}
\item For the \( \mathcal{N} = 2^* \) theory, where \( \epsilon_1 = \epsilon_2 = 0 \) and \( \epsilon_3 = - \epsilon_4 \);
\item For the massless \( \Omega \)--deformation of \( \mathcal{N} = 4 \) with \( \epsilon_1 = - \epsilon_2 \) and \( \epsilon_3 = \epsilon_4 = 0 \). 
\end{enumerate}
We start from the standard \D3--brane solution
\begin{align}
  \di s^2 &= H(r)^{-1/2} \di \vec x^2_{0 \dots 3} + H(r)^{1/2} \left( \di r^2 + r^2 \di \Omega_5^2 \right),\\
  F_4 &= \di H(r)^{-1} \wedge \di x^0 \wedge \ldots \wedge \di x^3 + 4 Q \ \omega_{S^5} \, ,
\end{align}
where \( H(r) = a + Q/ r^4 \). 
$r$ is the distance from the center of the D--brane and $Q$ is the D--brane charge. The coefficient $a$ is equal to zero at the horizon.

In the \( \mathcal{N} = 2^*  \) case, the lowest order deformation appears in the two-form fields:
\begin{align}
  B &= a V \wedge dx^8 + \frac{Q }{r^4} \left( V \wedge d x^8 + x^8 \omega \right), \\
   C_2 &= - \frac{Q}{r^4} \left( V \wedge d x^9 + x^9 \omega \right),
\end{align}
where \( 2 \omega  = \di V\). In the limit far away from the brane (corresponding to $Q=0$), the fluxtrap solution of flat space is recovered ($a V \wedge dx^8 $ being the first term of the $\epsilon$--expansion of the solution given in Eq.~\ref{eq:BB}). At the horizon ($a=0$) on the other hand, the form of the Polchinski-Strassler solution is recovered.

In the case of the \( \Omega \)--deformation of \( \mathcal{N}=4 \) \ac{sym} it is first of all necessary to analytically continue the undeformed solution or, equivalently, consider a solution of type \textsc{ii}* string theory~\cite{Hull:1998vg}. The undeformed is background is then $\mathrm{dS}_5 \times H^5$ and the deformation at first order is given by
\begin{align}
  B &= \left( V \wedge dx^8 -  \frac{Q}{Q + a r^4} x^8 \omega \right), \\
  C_2 &= - \frac{ Q }{Q + a r^4} x^8 \omega \,.
\end{align}

In both cases conformal invariance is broken. This corresponds to the presence of a non-trivial dilaton and \( C_0 \)--field in the near-horizon. Respectively
\begin{equation}
  \begin{cases}
    \Phi = - \frac{a V \cdot V}{2}  - \frac{Q \epsilon^2}{2} \frac{x_9^2 - x_8^2}{r^4}\\
   C_0 = Q\epsilon^2 \frac{x^8 x^9}{r^4}
 \end{cases}
\end{equation}
for \( \mathcal{N} = 2^* \) and
\begin{equation}
  \begin{cases}
    \Phi = - \frac{a V \cdot V}{2} + 6 \epsilon^2 \frac{Q}{r^2} + \dots \\
    C_0 = 3 \epsilon^2 \frac{Q}{r^2} + \dots
  \end{cases}
\end{equation}
for \( \Omega \)--deformed \( \mathcal{N}=4 \).

The metric deformation, which we expect to be of second order in \( \epsilon \) and issues related to the Myers' effect~\cite{Myers:1999ps} are currently under investigation.

\section{Conclusions}\label{sec:conc}

The fluxtrap background of string theory provides a transparent and algorithmic way of constructing supersymmetric gauge theories with both mass and Omega-type deformations. After reviewing the string theory background itself, we have discussed a number of explicit examples of two and four-dimensional gauge theories encoding the low energy effective description of the dynamics of D--branes in the fluxtrap background.

The fluxtrap approach can serve as a toolbox for the study of deformed supersymmetric gauge theories and their intimate relation to integrable models from a string theory perspective, a connection from which both fields can benefit greatly. It moreover provides a new route to gravity duals of deformed $\mathcal{N}=4$ theories. 

The fluxtrap construction is a starting point from which the wealth of existing results in the field of supersymmetric gauge theories which have emerged in recent years from different contexts can be meaningfully related and put onto a common ground.

\section*{Acknowledgments}

We would like to thank Ignatios Antoniadis, Marco Bill\`o, Marialuisa Frau, Valentina Forini, Francesco Fucito, Simeon Hellerman, Neil Lambert, Alberto Lerda, and Igor Pesando for enlightening discussions. We would moreover like to thank the Institute for Physics of the Humboldt University Berlin, the Simons Center for Physics and Geometry, and the Galileo Galilei Institute for Theoretical Physics for hospitality, and the \textsc{infn} for partial support during the completion of this work.

\printbibliography

\end{document}